\documentstyle[multicol,twoside,eqsecnum,epsf,aps,prb]{revtex}
\makeatletter

  \def\refrule{%
  \end{multicols}\widetext\vglue13pt %
  \hskip10.25pc\rule{20pc}{.1mm}\hfill %
  \vglue.14cm\begin{multicols}{2}\narrowtext %
  }
  \def\references{%
  \list{\@biblabel{\arabic{enumiv}}}%
  {\labelwidth\WidestRefLabelThusFar  \labelsep2pt %
  \leftmargin\labelwidth %
  \advance\leftmargin\labelsep %
  \ifdim\baselinestretch pt>1 pt %
  \parsep  4pt\relax %
  \else   %
  \parsep  0pt\relax %
  \fi
  \itemsep\parsep %
  \usecounter{enumiv}%
  \let\p@enumiv\@empty
  \def\theenumiv{\arabic{enumiv}}%
  }%
  \let\newblock\relax %
  \sloppy\clubpenalty4000\widowpenalty4000
  \sfcode`\.=1000\relax
  \ifpreprintsty\else\small\fi
  }
  \newfont{\rmtop}{cmr10 at 9pt}
 
  \def\AUTHORS{\rmtop  LO\"{I}C TURBAN AND FERENC IGL\'OI}
  \def\JOURNAL{PHYSICAL REVIEW B}
  \def\DATE{ 2002}
  \def\TITLE{\rmtop SURFACE-INDUCED DISORDER AND APERIODIC ...}
  \def\VOLUME{66}
  \def\NUMBER{}
  \def\PAGE{1}

  \def\ps@plain{%
  \gdef\@oddhead{\ifnum\thepage=\PAGE {\hbox to 2in{\rmtop\JOURNAL%
  \hfil}\hfil{{\rmtop VOLUME \VOLUME, NUMBER \NUMBER}}\hfil\hbox to %
  2in{\hfil{\rmtop\DATE}}}\else{\hbox to %
  1in{$\underline{\hbox{\rmtop\VOLUME}}$\hfil}\hfil\TITLE\hfil\hbox to %
  1in{\hfil\rmtop\thepage}}\fi}% 
  \gdef\@evenhead{\hbox to % 
  1in{\rmtop\thepage\hfil}\hfil\AUTHORS%  
  \hfil\hbox to 1in{\hfil$\underline{\hbox{\rmtop\VOLUME}}$}}% 
  \gdef\@oddfoot{\ifnum\thepage=\PAGE{\hbox to %
  2.5in{\hfil}\hfil$\ \ \underline{\hbox{\rmtop\VOLUME}}\ \ \ \ \ \ %
  ${\rmtop\thepage}%  
  \hfil\hbox to 2.5in{\hfil Typeset using REV\TeX}}\else{}\fi}% 
  \gdef\@evenfoot{{}}% 
  }%

  \makeatother

\begin{document}

\title{Surface-induced disorder and aperiodic 
perturbations at first-order transitions}

\author{Lo\"{\i}c Turban}

\address{Laboratoire de Physique des Mat\'eriaux, Universit\'e 
Henri Poincar\'e
(Nancy 1), B.P. 239,\\ F-54506 Vand\oe uvre l\`es Nancy cedex, 
France}

\author{Ferenc Igl\'oi}

\address{Research Institute for Solid State Physics and Optics, 
P.O. Box 49,
H-1525 Budapest,  Hungary\\
and Institute for Theoretical Physics,
Szeged University, H-6720 Szeged, Hungary}

\date{Received 19 October 2001; revised manuscript received 4 April 2002}

\maketitle

\begin{abstract}
In systems displaying a bulk first-order transition the order 
parameter may vanish
continuously at a free surface, a phenomenon which is called 
{\it 
surface-induced 
disorder}. In the presence of surface-induced disorder the 
correlation lengths, parallel 
and perpendicular to the surface, diverge at the bulk 
transition 
point.
In this way the surface induces an anisotropic power-law 
singular 
behavior 
for some bulk quantities. For example, in a finite system of 
transverse 
linear size $L$, the response functions diverge as 
$L^{(d-1)z+1}$, where 
$d$ is the dimension of the system and $z$ is the anisotropy 
exponent. 
We present a general scaling picture for this {\it anisotropic 
discontinuity 
fixed point}. Our phenomenological results are confronted with 
analytical and
numerical calculations on the two-dimensional $q$-state Potts model in 
the large-$q$ limit. The scaling results are demonstrated to 
apply also for 
the same model with a layered, Fibonacci-type modulation of the 
couplings for which 
the anisotropy exponent is a continuous function of the 
strength 
of 
the quasiperiodic perturbation.
\end{abstract}

\pacs{05.50.+q, 64.60.Cn, 68.35.Rh, 61.44.-n} 

\begin{multicols}{2}
\narrowtext

\section{Introduction}

At a first-order phase transition several physical quantities, 
such as
the order parameter and the energy density, have a 
discontinuous 
behavior 
in the bulk and the correlation lengths remain generally 
finite. 
In spite of the absence of a diverging ``true" length scale in 
the problem, 
a successful scaling theory for first-order transitions
has been developed,\cite{nn,fb} with scaling exponents such as
\begin{equation}
\beta_D=0,\quad\alpha_D=\gamma_D=1,\quad\delta_D=\infty\;,
\label{D_exp}
\end{equation} 
which are compatible with discontinuities. The divergent 
length scale appearing in the scaling theory is associated with 
the 
persistence length of the coexisting phases,\cite{fb} 
behaving as $\vert t\vert^{-\nu_D}$ ($\vert h\vert^{-\nu_D^h}$)
for a temperature (field) deviation from the transition point,
with critical exponents
\begin{equation}
\nu_D=1/d\;,\qquad\nu_D^h=1/d\;,
\label{nu_D}
\end{equation} 
where $d$ denotes the dimension of the system. 
In the renormalization group analysis of specific models,
first-order transitions have been identified,\cite{nn,nbrs,sp}
which are controlled by a so-called discontinuity fixed point 
(DFP) 
with thermal and magnetic scaling dimensions
$y_t=\ln \Lambda_t/\ln b=d$ and $y_h=\ln \Lambda_h/ \ln b=d$.
These dimensions are compatible with the value of $\nu_D$ and 
$\nu_D^h$ in Eq.~(\ref{nu_D}) whereas other scaling exponents 
in Eq.~(\ref{D_exp}) follow from scaling and hyperscaling 
relations.

For a system with a free surface at a bulk first-order 
transition 
two diverging lengths can be identified, $\xi_{\parallel}$ 
associated with the correlations in the direction 
parallel to the surface and $\xi_{\perp}$ in the perpendicular 
direction.
Due to missing bonds, the surface is more weakly coupled than
the bulk. As a consequence, the surface order parameter 
vanishes 
continuously at the bulk transition point. This phenomenon,  
called {\it surface-induced disorder}\cite{lip1} (SID), is 
closely 
related to wetting phenomena.\cite{diet} It has been studied 
using mean-field theory\cite{lip2,kg} and Monte Carlo
simulations\cite{gk} and has been observed in some experimental 
systems.\cite{do}
For the two-dimensional (2D) Potts model exact results have 
been 
obtained in the large-$q$ limit and the universality
of the critical and tricritical exponents in the bulk 
first-order 
transition regime, $q>4$, has been accurately demonstrated by 
density matrix renormalization calculations.\cite{ic}

The presence of diverging length scales in the problem 
suggests a possible scaling behavior for 
singular quantities, both at the surface and in the bulk 
of the system. Here we notice two peculiarities 
concerning the bulk scaling relations. First, in the geometry 
with a free surface there is an inherent anisotropy in the 
problem
since the two length scales are related by
\begin{equation}
\xi_{\parallel} \sim \xi_{\perp}^z\;,
\label{def_z}
\end{equation} 
with an anisotropy exponent $z$ which is generally greater 
than~1. Consequently, scaling for SID is strongly anisotropic 
in contrast to the original DFP theory, which was developed
for isotropic scaling in the bulk.\cite{fb} 
Our second remark concerns the behavior of the system as 
the thermodynamic limit is approached. In a system of linear 
size $L$, the disordering effect of the surface in the bulk 
is limited to a temperature range $\Delta t\sim 
L^{-1/\nu_{\perp}}$
around the critical temperature, where $\nu_{\perp}$ is 
the perpendicular correlation length exponent [see below 
in Eq.~(\ref{xi_perp})]. In the thermodynamic limit, when 
$L\to\infty$, the SID is present in the bulk
just at the transition point, when the scaling functions 
are singular, having either a jump or $\delta$-function-like 
singularities. In a finite system, however, the scaling 
functions behave regularly, so that one has a ``true" 
finite-size scaling behavior, which is measurable,
for the specific heat or the susceptibility 
at the transition point.

Another problem of theoretical interest at a first-order 
transition point accompanied by SID is the influence of 
different types of perturbations, such as localized or 
extended defects,\cite{ipt} quenched disorder,\cite{har} 
etc., on the bulk and surface transitions. Here we mainly 
consider the effect of a quasiperiodic or, more generally, 
aperiodic modulation of the couplings in the bulk. The 
influence of such perturbations on order-disorder
phase transitions has been studied in detail for
the Ising model,\cite{qpim} in particular 
in 2D with a layered structure (i.e., when
the aperiodic perturbation has a 1D variation) 
and in the related quantum Ising spin chains. Among others, 
for marginal aperiodic sequences, coupling-dependent 
anisotropic critical behavior has been 
found.\cite{it,itksz,hgb}

At a first-order transition point, such as for the 
2D Potts model with $q>4$,\cite{baxt,wu} the 
relevance or irrelevance of the effect of aperiodic 
perturbations is unclear. One can use an analogy 
with the rigorous result of Aizenman and Wehr,\cite{aw} 
stating that in 2D a first-order
transition always softens into a second-order one, 
due to quenched disorder. Such a  softening of the 
transition was indeed observed in a Monte Carlo study 
of the 2D Potts model with $q>4$ for some types of 
layered aperiodic perturbations.\cite{bcb} However, 
due to computational difficulties, quantitative results 
about the phase transitions, such as critical exponents, 
could not be obtained.

In the present paper we use another approach to study 
the effect of aperiodic perturbations at a first-order 
transition point and to obtain quantitative 
results. For this purpose we consider the 2D Potts 
model\cite{wu} on layered lattices and take the 
extreme anisotropic limit,\cite{kog} when the logarithm 
of the transfer matrix of the model corresponds to the 
Hamiltonian of an aperiodic quantum Potts chain. 
Then we consider the large-$q$ limit of the model, 
for which a consistent $1/\sqrt{q}$ expansion is defined 
and the leading term can be treated by analytical 
or very efficient numerical methods. In analogy
with the results on the marginal quantum Ising chains, 
one expects that with marginal perturbations 
the critical exponents, in particular the anisotropy 
exponent, become coupling dependent. Indeed, with a 
Fibonacci-type modulation of the couplings, which 
is related to the 2D Penrose quasilattice, we have 
found continuously varying anisotropic scaling. 
In this way we have obtained a more general testing 
ground for the scaling relations in the presence of SID.

The structure of the paper is the following. 
In Sec.~II we present the phenomenological scaling 
theory for SID both in the bulk and at the surface.
In Sec.~III we introduce the quantum Potts model 
and its representation in the large-$q$ limit.
In Sec.~IV solutions of the model on the regular lattice 
and on the Fibonacci quasilattice are confronted with 
predictions of the SID scaling picture. The results 
are discussed in Sec.~V.

\section{Scaling behavior with surface-induced disorder}

We consider a $d$-dimensional (magnetic) system in the 
slab geometry with two surfaces at a large
finite distance $L$ apart, whereas the extent of 
the system is infinite in the other $(d-1)$ dimensions,
parallel to the surface. The system 
displays a bulk first-order transition and has
$q$ degenerate ordered phases. The state of the 
system on the surface at $l=L$ is fixed into one of these
phases (by applying a local magnetic field in the 
appropriate direction) whereas the other
surface at $l=1$ is free. The temperature $t$ is measured
relative to the transition point and
a bulk ordering (magnetic) field of strength 
$h$ is applied. Due to the disordering effect of missing 
bonds, the order parameter profile $m_l$ is monotonically 
decreasing towards the free surface. 
The surface order parameter $m_1$ may behave in 
two qualitatively different ways as the
temperature is increased to the transition point. 
Either $m_1$, like the bulk order parameter $m_b=m_{L/2}$, 
has a finite jump at $t=0$ (first-order surface transition) 
or---and this is the case considered below---$m_1$ vanishes 
continuously as
\begin{equation}
m_1 \sim (-t)^{\beta_1},\qquad t<0\;,
\label{beta1}
\end{equation} 
with a surface magnetization exponent $\beta_1$ (second-order 
surface transition). In this case the extent of the surface 
region $\xi_{\perp}$, where $m_l\sim m_1$ ($1<l<\xi_{\perp}$), 
is diverging at the transition point as
\begin{equation}
\xi_{\perp} \sim (-t)^{-\nu_{\perp}}\;.
\label{xi_perp}
\end{equation} 
As mentioned in the Introduction, this phenomenon
associated with the long-distance disordering effect
of a free surface is called SID.

In the region of SID, correlations between spins in the 
direction parallel to the free surface decay exponentionally, 
which defines a new correlation length $\xi_{\parallel}$, 
with an asymptotic temperature dependence
\begin{equation}
\xi_{\parallel} \sim (-t)^{-\nu_{\parallel}}\;.
\label{xi_parallel}
\end{equation} 
According to Eq.~(\ref{def_z}) the correlation-length 
exponents satisfy the relation $\nu_{\parallel}=z 
\nu_{\perp}$. Going through the transition point 
by varying $h$ at $t=0$ leads to a similar behavior with 
$-t$ replaced by $h$ in Eqs.~(\ref{xi_perp}) and 
(\ref{xi_parallel}), while the corresponding critical exponents
are $\nu_{\perp}^h$ and $\nu_{\parallel}^h$, 
respectively.

Now using the scaling hypothesis for strongly ani\-so\-tro\-pic 
systems\cite{anis} we can write the following relation 
for the singular part of the bulk free-energy density:
\begin{equation}
f(t,h,L)=b^{-[(d-1)z+1]} f(t b^{1/\nu_{\perp}}, h 
b^{1/\nu_{\perp}^h},L/b)\;,
\label{fe_scaling}
\end{equation} 
when lengths are rescaled by a factor $b>1$. By 
derivation one obtains a similar
scaling relation for the bulk order parameter,
\begin{eqnarray}
&&m_b(t,h,L)=\frac{\partial f}{\partial h}\nonumber\\
&&\ \ \ \ \ =b^{-[(d-1)z+1]+1/\nu_{\perp}^h} m_b(t 
b^{1/\nu_{\perp}}, 
h b^{1/\nu_{\perp}^h},L/b)\;,
\label{mb_scaling}
\end{eqnarray} 
and for the excess internal energy per degree of freedom,
\begin{eqnarray}
&&\Delta u(t,h,L)=u(t,h,L)-u(0,0,\infty)
=\frac{\partial f}{\partial t}\nonumber\\
&&\ \ \ \ \ =b^{-[(d-1)z+1]+1/\nu_{\perp}}\Delta u(t 
b^{1/\nu_{\perp}}, 
h b^{1/\nu_{\perp}^h},L/b)\;.
\label{u_scaling}
\end{eqnarray} 
Since the first-order transition is accompanied by a 
magnetization jump and a finite latent heat, both  
$m_b(t,h,L)$ and $\Delta u(t,h,L)$ approach a 
scale-independent, finite limit
when $L\to\infty$. Thus Eqs.~(\ref{mb_scaling}) and 
(\ref{u_scaling}) lead to the scaling relations
\begin{equation}
\nu_{\perp}=\nu_{\perp}^h=\frac{1}{(d-1)z+1}\;.
\label{nu_exp}
\end{equation} 
One may notice that in the case of isotropic scaling, i.e., 
with 
$z=1$, we recover the result of the DFP in Eq.~(\ref{nu_D}).

For the susceptibility $\chi=\partial^2 
f/\partial h^2$ and the specific heat
$c_v=\partial^2 f/\partial t^2$, using the scaling relation 
in Eq.~(\ref{fe_scaling}), we obtain
\begin{eqnarray}
\chi(t,h=0,L)&=&t^{-\gamma_D} \widetilde{\chi}_1(L 
t^{\nu_{\perp}})
=L^{(d-1)z+1}\widetilde{\chi}_2(L t^{\nu_{\perp}})
\nonumber\\
c_v(t,h=0,L)&=&t^{-\alpha_D} \widetilde{c_v}_1(L 
t^{\nu_{\perp}})
=L^{(d-1)z+1}\widetilde{c_v}_2(L t^{\nu_{\perp}})\;,\nonumber\\
\label{chi_cv_scaling}
\end{eqnarray}
with the DFP exponents $\gamma_D$ and $\alpha_D$ 
in Eq.~(\ref{D_exp}). The scaling functions behave as
$\widetilde{\chi}_1(y)\sim\widetilde{c_v}_1(y)\sim 
y^{(d-1)z+1}$ 
and $\widetilde{\chi}_2(y)\sim\widetilde{c_v}_2(y)
={\rm const}$ as $y\to 0$.
Thus, {\it in a finite system}, when approaching the transition 
point, one can observe the development of a singularity with 
the
DFP exponents in Eq.~(\ref{D_exp}). However, in the 
thermodynamic limit $L\to\infty$, this singularity
transforms into a $\delta$ function as it should at 
a first-order transition point. Indeed the temperature
range where the surface can influence the behavior 
of the bulk is limited to $\Delta t\sim L^{-1/\nu_{\perp}}$,
so that $\Delta t\chi(t=0,h=0,L)\approx 1$, and in 
this way $\chi(t,h=0,L)$---and similarly $c_v(t,h=0,L)$---is 
a representation of the $\delta$ function as 
$L\to\infty$. On the other hand,
{\it at the transition point} the response functions 
show a power-law finite-size dependence, with the same critical 
exponent
$(d-1)z+1$ as given in
Eq.~(\ref{chi_cv_scaling}) and which should be measurable 
in finite samples.

For the correlation lengths one can similarly write 
the scaling relations
\begin{eqnarray}
\xi_{\perp}(t,h,L)&=& b\,\xi_{\perp}(t b^{1/\nu_{\perp}}, 
h b^{1/\nu_{\perp}^h},L/b)\;,
\nonumber\\
\xi_{\parallel}(t,h,L)&=&b^z\, \xi_{\parallel}(t 
b^{1/\nu_{\perp}}, 
h b^{1/\nu_{\perp}^h},L/b)\;,
\label{xi_scaling}
\end{eqnarray}
from which the temperature (field) dependence in 
Eqs.~(\ref{xi_perp}) and (\ref{xi_parallel}) follows 
by taking the length scale as $b=t^{-\nu_{\perp}}$ 
($b=h^{-\nu_{\perp}^h}$).

We close this section with a similar analysis of the 
singular behavior at the surface. For the singular part 
of the surface free energy per 
degree of freedom we write the scaling relation
\begin{eqnarray}
&&f_1(t,h,h_1,L)\nonumber\\
&&\ \ =b^{-(d-1)z} f_1(tb^{1/\nu_{\perp}}, 
hb^{1/\nu_{\perp}^h},h_1b^{(d-1)z-x_1},L/b),
\label{f1_scaling}
\end{eqnarray} 
where a surface ordering field $h_1$ with anomalous scaling
dimension $x_1$ has been included. For the surface order 
parameter $m_1=\partial f_1/\partial h_1$, 
Eq.~(\ref{f1_scaling}) 
leads to the following scaling behavior:
\begin{equation}
m_1(L)\sim L^{-x_1},\qquad m_1(t)\sim t^{x_1\nu_{\perp}}\;.
\label{m1_scaling}
\end{equation} 
Comparing Eqs.~(\ref{xi_perp}) and~(\ref{m1_scaling}) 
we obtain
\begin{equation}
\beta_1=x_1\nu_{\perp}\;.
\label{beta_1}
\end{equation} 
Taking other derivatives of $f_1$ one obtains 
similar scaling relations for the surface
susceptibility $\chi_{1,1}=\partial^2 f_1/\partial h_1^2$, 
for the excess surface order parameter 
$m_s=\partial f_1/\partial h$, etc.

To conclude this section, let us remark that at a first-order 
transition point which is accompanied by
SID, the bulk singularities 
in finite systems involve the anisotropy
exponent $z$ and the dimension $d$ of the system 
whereas the surface singularities involve
a new exponent $x_1$. In the following section 
our scaling assumptions are confronted with
analytical and numerical results on specific models.

\section{The two-dimensional Potts model in the large-{\boldmath\lowercase{q}} limit}

We start with the Hamiltonian of the classical $q$-state 
Potts model on the square lattice,\cite{wu}
\begin{eqnarray}
-\beta H&=&\sum_{k=-\infty}^{\infty}\sum_{l=1}^{L} 
K_1(l)\left[\delta(s_{l,k+1}-s_{l,k})-1\right]\nonumber\\
&&\  
+\sum_{k=-\infty}^{\infty}\sum_{l=1}^{L-1} 
K_2(l)\left[\delta(s_{l+1,k}-s_{l,k})-\frac{1}{q}\right]\;,
\label{hamilton}
\end{eqnarray} 
where $s_{l,k}=1,2,\dots,q$ is a $q$-state Potts variable 
defined modulo $q$ at site ($l,k$). The vertical [horizontal] 
couplings $K_1(l)$ [$K_2(l)$] are ferromagnetic and may 
depend on the horizontal position $l$. Constant terms have 
been added in order to simplify some expressions in the following.
The transfer operator $T$ has matrix elements
\begin{eqnarray}
\langle s\vert T\vert s'\rangle &=& 
\langle s\vert U\vert s\rangle 
\langle s\vert V\vert s'\rangle\;,\nonumber\\ 
\langle s\vert U\vert s\rangle &=& 
\exp\left\{\sum_{l=1}^{L-1} K_2(l)
\left[\delta(s_{l+1}-s_l)-\frac{1}{q}\right]\right\}\;,
\nonumber\\ 
\langle s\vert V\vert s'\rangle &=&
\exp\left\{\sum_{l=1}^L
K_1(l)\left[\delta({s'}_l-s_l)-1\right] \right\}\;,
\label{tmatrix}
\end{eqnarray} 
between the Potts states $\vert s\rangle$ and $\vert s'\rangle$ 
associated with two successive rows $k$ and $k+1$. 
Let us associate with each site $l$ along a row the Potts 
operators $\Omega_l$ and $M_l$ such that\cite{mit}
\begin{eqnarray}
\Omega_l\vert s_l\rangle &=&
\exp\left(i\frac{2\pi}{q}s_l\right)\vert s_l\rangle\;,\quad
\Omega^\dagger_l=\Omega^{-1}_l\;,\nonumber\\
M_l\vert s_l\rangle &=& \vert s_l+1\rangle\;,\quad
M^\dagger_l=M^{-1}_l\;.
\label{oper}
\end{eqnarray}
Thus $\Omega_l$ is diagonal in the basis of the Potts states 
whereas 
$M_l$ is a ladder operator. On the same site, they satisfy the 
commutation rules
\begin{eqnarray}
M_l\Omega_l &=& \exp\left(-i\frac{2\pi}{q}\right)\Omega_l 
M_l\;,\nonumber\\
M^\dagger_l\Omega_l &=& 
\exp\left(i\frac{2\pi}{q}\right)\Omega_l 
M^\dagger_l\;,
\label{commu}
\end{eqnarray}
and commute on different sites. Making use of the identity
\begin{equation}
\delta(s_{l+1}-s_l)=\frac{1}{q}\sum_{p=0}^{q-1}
\exp\left[i\frac{2p\pi}{q}(s_{l+1}-s_l)\right]\;,
\label{ident}
\end{equation}
the diagonal operator $U$ can be written as
\begin{equation}
U=\exp\left\{\sum_{l=1}^{L-1}\frac{K_2(l)}{q}
\sum_{p=1}^{q-1}(\Omega_l\Omega^\dagger_{l+1})^p\right\}\;.
\label{u}
\end{equation}
In Eq.~(\ref{tmatrix}) the diagonal matrix element 
$\langle s\vert V\vert s\rangle$ is equal to 1. When the two 
states 
$\vert s\rangle$ and $\vert s'\rangle$ differ only on site $l$, 
we have   
$\langle s\vert V\vert s'\rangle=\exp[-K_1(l)]$. Thus one may 
write
the nondiagonal part of the transfer operator as
\begin{equation}
V=1+\sum_{l=1}^L\exp[-K_1(l)]\sum_{p=1}^{q-1}M^p_l+\cdots,
\label{v}
\end{equation}
where the ellipsis stands for higher-order terms involving products 
of 
exponentials coming from different sites. 

In the extreme anisotropic limit\cite{kog} where 
$K_1(l)\to\infty$, 
$K_2(l)\to 0$, and $\vartheta\to 0$ in such a way that the 
ratios
\begin{equation}
J_l=\frac{K_2(l)}{\vartheta}\;,\qquad 
\Gamma_l=q\frac{\exp[-K_1(l)]}{\vartheta}
\label{couplings}
\end{equation}
remain finite, the operators in Eqs.~(\ref{u}) and~(\ref{v}) 
are now 
given by
\begin{eqnarray}
U &=& 1+\vartheta\sum_{l=1}^{L-1}\frac{J_l}{q}
\sum_{p=1}^{q-1}(\Omega_l 
\Omega^\dagger_{l+1})^p+O(\vartheta^2)\;,
\nonumber\\
V &=& 
1+\vartheta\sum_{l=1}^L\frac{\Gamma_l}{q}\sum_{p=1}^{q-1}M^p_l
+O(\vartheta^2)\;.
\label{uv}
\end{eqnarray}
To first order in $\vartheta$, $U$ and $V$ commute and the 
transfer operator 
can be rewritten as the evolution operator 
$T=\exp(-\vartheta{\cal H})$ for 
the infinitesimal Euclidian time step $\vartheta$. According to 
Eqs.~(\ref{tmatrix}) and~(\ref{uv}), the quantum Potts 
Hamiltonian
takes the following form:
\begin{equation}
{\cal H}=-\sum_{l=1}^{L-1}\frac{J_l}{q}
\sum_{p=1}^{q-1}(\Omega_l \Omega^\dagger_{l+1})^p
-\sum_{l=1}^L\frac{\Gamma_l}{q}\sum_{p=1}^{q-1}M^p_l\;.
\label{hamiltonop}
\end{equation} 
Dual operators which satisfy the Potts algebra in 
Eq.~(\ref{commu}) are defined 
as follows:
\begin{eqnarray}
\widehat{M}_l &=& \Omega_{l-1}\Omega^\dagger_l\qquad 
(l=2,L)\;,\nonumber\\ 
\widehat{M}_1 &=& \Omega^\dagger_1\nonumber\\
\widehat{\Omega}_l &=& 
\prod_{i=0}^{L-l-1}M_{l+i}M^\dagger_{l+i+1}
\qquad (l=1,L-1)\;,\nonumber\\
\widehat{\Omega}_L &=& M_L\;.
\label{duop}
\end{eqnarray}
Under the duality transformation the Hamiltionain in 
Eq.~(\ref{hamiltonop}) is 
changed into
\begin{equation}
{\cal H}\!=\!-\!\sum_{l=2}^L\!\frac{J_{l-1}}{q}\!
\sum_{p=1}^{q-1}\!\widehat{M}^p_l\!
-\!\sum_{l=1}^{L-1}\!\frac{\Gamma_l}{q}\!\sum_{p=1}^{q-1}
\!(\widehat{\Omega}_l \widehat{\Omega}^\dagger_{l+1})^p\!
-\!\frac{\Gamma_L}{q}\sum_{p=1}^{q-1}\!\widehat{\Omega}^p_L\,.
\label{hamiltondu}
\end{equation} 
Having fixed boundary conditions with $s_L=1$ amounts to have a 
vanishing transverse 
field $\Gamma_L$, which eliminates the last anomalous term in 
the dual 
Hamiltonian. 
The coupling $J_{l-1}$ is transformed into a transverse field 
acting on the 
Potts state at $l$ while the transverse field $\Gamma_l$ 
becomes a coupling 
between Potts states at $l$ and $l+1$. It follows that the 
original system with 
free-boundary conditions at $l=1$ and fixed-boundary conditions 
at $l=L$ is 
transformed into a dual system with fixed-boundary conditions 
at $l=1$ and free-boundary conditions at $l=L$. The Hamiltonian 
is strictly self-dual when 
$\Gamma_l=J_l$ either for homogeneous couplings $\Gamma=J$ or, 
with 
inhomogeneous couplings, when the sequence is symmetric. Then 
the  self-duality 
point $J_l=\Gamma_l$ corresponds to the critical point. 

In order to study the effect of a homogeneous longitudinal
magnetic field, the Hamiltonian in Eq.~(\ref{hamiltonop}) 
can be completed by the diagonal field term
$-(h/\sqrt{q})\sum_{l=1}^{L-1}[q\delta(s_l-1)-1]/(q-1)$.

In the calculations we consider the behavior of the system 
in the vicinity of the self-duality point,
taking the parametrization
\begin{equation}
J_l=\Gamma_l\left(1-\frac{t}{\sqrt{q}}\right)\;,
\label{param}
\end{equation} 
where $t$ plays the same role as the 
temperature in the classical model: for $t>0$
($t<0$) we are in the bulk paramagnetic (ferromagnetic) 
phase with a vanishing (nonvanishing)
order parameter. The local order 
parameter, i.e., the magnetization profile
at site $l$, is defined as
\begin{equation}
m_l=\frac{q\langle0\vert\delta(s_l-1)\vert
0\rangle -1}{q-1}\;,
\label{magn}
\end{equation} 
where $\vert0\rangle$ is the ground-state 
eigenvector of ${\cal H}$. On the other hand, 
the free-energy density
of the 2D classical model is proportional to 
the ground-state energy per site, 
$e_0=\langle0\vert{\cal H}\vert0\rangle$.\cite{kog}

In the large-$q$ limit one may use a systematic 
expansion in powers of $1/\sqrt{q}$. To leading
order the following eigenfunctions are degenerate:
\end{multicols}
\widetext
\noindent\rule{20.5pc}{.1mm}\rule{.1mm}{2mm}\hfill
\begin{eqnarray}
\vert\psi_1 \rangle&=&\vert11 \dots 11 \rangle, \nonumber \\
\vert\psi_2\rangle&=&\frac{1}{q^{1/2}}(\vert 21 \dots 11
\rangle+\vert 31 \dots 11\rangle+\dots +\vert q1 \dots 
11\rangle), 
\nonumber \\
\vert\psi_3\rangle&=&\frac{1}{q}[\vert 321\dots11\rangle
+\vert 421 \dots 11\rangle+\dots +\vert q(q-1)1 \dots 
11\rangle], 
\nonumber \\
\vert\psi_4\rangle&=&\frac{1}{q^{3/2}}[\vert 4321\dots 
11\rangle
+\vert 5321\dots 11\rangle+\dots +\vert q(q-1)(q-2)1 \dots 
11\rangle], \nonumber \\
&\vdots& \nonumber \\
\vert\psi_L\rangle&=&\frac{1}{q^{(L-1)/2}}[\vert L 
\dots 4321\rangle+\dots +\vert q(q-1)(q-L)\dots 
(q-L+2)1\rangle],
\label{LL}
\end{eqnarray}
all having the same diagonal contribution
$\langle\psi_i\vert{\cal H}\vert\psi_i\rangle=-\sum_{l=1}^{L-1} 
J_l=-\sum_{l=1}^{L-1}\Gamma_l$, where we made
use of Eq.~(\ref{param}). 

This degeneracy is lifted to the next order in $1/\sqrt{q}$ 
and the secular problem is governed by the operator
\begin{equation}
{\widetilde {\cal H}} =-\frac{1}{\sqrt{q}} \left(
\matrix{
0&\Gamma_1&&&&0\cr
\Gamma_1&t\Gamma_1-h&\Gamma_2&&&\cr
&\Gamma_2&\sum_{l=1}^2 (t\Gamma_l-h)&\Gamma_3&&\cr
&&\Gamma_3& \sum_{l=1}^3(t\Gamma_l-h)&\ddots&\cr
&&&\ddots&\ddots&\Gamma_{L-1}\cr
0&&&&\Gamma_{L-1}&\sum_{l=1}^{L-1}(t\Gamma_l -h)\cr}
\right)\;,
\label{htilde}
\end{equation} 
\hfill\rule[-2mm]{.1mm}{2mm}\rule{20.5pc}{.1mm}
\begin{multicols}{2} 
\narrowtext
\noindent up to a constant term 
$-\sum_{l=1}^{L-1}\Gamma_l-(L-1)h/\sqrt{q}$.
One may notice that for the homogeneous problem with 
$\Gamma_l=\Gamma$, the temperature and longitudinal 
field play the same role since we have 
the symmetry $t\Gamma\leftrightarrow -h$.
The ground-state eigenvector of ${\cal H}$ is given 
by the linear combination
\begin{equation}
\vert0\rangle=\sum_{i=1}^L v_i \vert\psi_i\rangle\;,
\end{equation} 
where $v_i$ are the components of the leading 
eigenvector of $-\widetilde {\cal H}$ in
Eq.~(\ref{htilde}). Making use of Eqs.~(\ref{magn}) 
and~(\ref{LL})
we obtain the magnetization profile as
\begin{equation}
m_l=\sum_{i=1}^l \vert v_i\vert^2\;,
\label{magn_P}
\end{equation} 
in the large-$q$ limit. In particular the surface 
magnetization is given by $m_1=\vert v_1\vert^2$  
and the bulk magnetization is defined as $m_b=m_{L/2}$.

The perpendicular correlation length 
$\xi_{\perp}$ gives the width of the interface region and, 
in the ordered phase, it can be 
defined through the relation
\begin{equation}
m_l\vert_{l=\xi_{\perp}}=\frac{1}{2}\;.
\label{xi_perp_P}
\end{equation} 
On the other hand, the parallel correlation length 
$\xi_{\parallel}$ 
of the 2D classical model is related to the imaginary time 
spin-spin correlation function $G_l(\tau)$ of the quantum 
model.
The long-time behavior of $G_l(\tau)$ has the asymptotic form 
\begin{equation}
G_l(\tau)=\langle0\vert\delta[s_l(0)-s_l(\tau)]\vert
0\rangle\sim\exp(-\tau\Delta)\;,
\label{auto}
\end{equation} 
where $\Delta$ is the gap at the top of the spectrum of 
$-\widetilde{\cal H}$. Since $\Delta\sim1/\xi_{\parallel}$,
one obtains information  about the behavior of the correlations 
in the 2D classical system from the scaling properties of the 
spectrum 
of the Hamiltonian.\cite{kog}

\section{Surface-induced disorder in regular and quasiperiodic 
lattices}

\subsection{Solution on the regular lattice}

For homogeneous interactions, i.e., when $\Gamma_l=\Gamma$, 
the complete solution of the Hamiltonian in Eq.~(\ref{htilde}) 
has been given in Ref.~\onlinecite{ic}. 
The critical exponents associated with the correlations 
parallel 
and perpendicular to the surface are given by
\begin{equation}
\nu_{\perp}=\nu_{\perp}^h=\frac{1}{3}\;,\qquad 
\nu_{\parallel}=\nu_{\parallel}^h=\frac{2}{3}\;,\qquad z=2\;.
\label{exp_hom}
\end{equation} 
%

%%%%%%%%%%%%%%%%% FIGURE %%%%%%%%%%%%%%%%%%%%%%%%%
\begin{figure}[tbh]
\epsfxsize=7truecm
\begin{center}
\mbox{\epsfbox{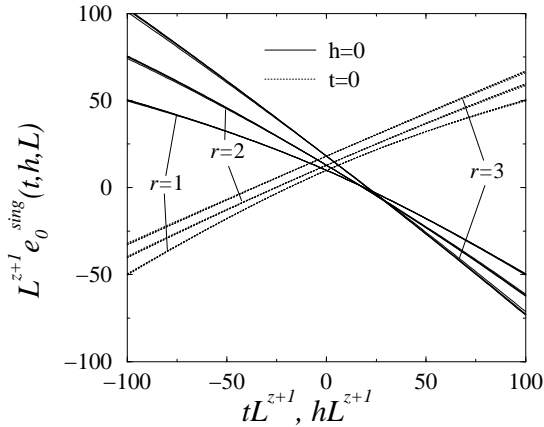}}
\end{center}
\caption{\label{fig1} Scaling plot of the singular part 
of the bulk energy density of the quantum Potts chain 
in the large-$q$ limit for the regular ($r=1$) and
the Fibonacci lattice ($r=2,3$). The lines for finite systems 
with sizes $L=89,~233,~610$, and $1597$ are practically 
indistinguishable.
} 
\end{figure}
%%%%%%%%%%%%%%%%%%%%%%%%%%%%%%%%%%%%%%%%%%%%%%%%%%

\noindent For the surface magnetization one obtains
\begin{equation}
x_1=3\;,\qquad \beta_1=1\;.
\label{exp_hom_surf}
\end{equation} 
First, one may notice that conventional scaling relations 
in Eq.~(\ref{beta_1}) and below Eq.~(\ref{xi_parallel}) 
are satisfied. More importantly, the relations in 
Eq.~(\ref{nu_exp}), which are a consequence of the 
discontinuous behavior in the bulk, are also satisfied.

\begin{figure}[tbh]
\epsfxsize=7truecm
\begin{center}
\mbox{\epsfbox{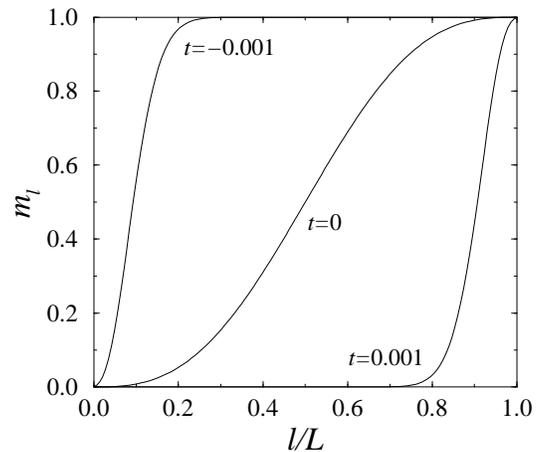}}
\end{center}
\caption{\label{fig2} Magnetization profiles of the 
homogeneous system with size $L=144$ at
different temperatures for free- and fixed-boundary conditions. 
The critical profile ($t=0$) is the analytical result of 
Eq.~(\protect\ref{prof}). The interface for $t<0$ ($t>0$) 
is localized at the free (fixed) boundary.
}
\end{figure}

Let us now consider the scaling of the free-energy density 
in Eq.~(\ref{fe_scaling}). For the quantum chain 
it corresponds to the singular part of the ground-state 
energy density, $e_0^{\rm 
sing}(t,h,L)=e_0(t,h,L)-e_0(0,0,\infty)$, 
with the following finite-size scaling behavior:\cite{rem}
\begin{equation}
e_0^{\rm 
sing}(t,h,L)=L^{-(z+1)}\widetilde{e}_0(tL^{z+1},hL^{z+1})\;.
\label{e0_scaling}
\end{equation} 
According to analytical results [see Eq.~(26) in 
Ref.~\onlinecite{ic}] 
this relation is indeed satisfied for the homogeneous system 
with
$z=2$. The numerical results for the scaling functions are 
shown in Fig.~\ref{fig1}. They agree with the result of a 
perturbation expansion 
$\widetilde{e}_0(y,0)=\pi^2-y/2+O(y^3)$ to linear order.
The scaling form of the magnetization profile in 
Eq.~(\ref{mb_scaling}) is given by
\begin{equation}
m_l(t,h,L)=m_{l/L}(tL^{z+1},hL^{z+1})\;.
\label{ml_scaling}
\end{equation} 
It is presented in Fig.~\ref{fig2} for $h=0$ and $t<0$, 
$t>0$, and $t=0$. In the latter case the
asymptotic scaling form is given by
\begin{equation}
m_{l/L}(0,0)=\frac{l}{L}-\frac{1}{2\pi}
\sin\left(\frac{2\pi l}{L}\right)\;.
\label{prof}
\end{equation} 
%
%%%%%%%%%%%%%%%%% FIGURE %%%%%%%%%%%%%%%%%%%%%%%%%
%%%%%%%%%%%%%%%%%%%%%%%%%%%%%%%%%%%%%%%%%%%%%%%%%%
\noindent In the ordered regime one can identify the interface 
region with width $\xi_{\perp}\sim t^{-1/(1+z)}$
near the surface at $l=1$. In the high-temperature phase 
the interface has a width $\overline{\xi}_{\perp}$. It is 
localized near the ordered surface at $l=L$. The two 
widths are related through $\overline{\xi}_{\perp}(t)=
\xi_{\perp}(-t)$, $t>0$.
%%%%%%%%%%%%%%%%% FIGURE %%%%%%%%%%%%%%%%%%%%%%%%%
\begin{figure}[bht]
\epsfxsize=7truecm
\begin{center}
\mbox{\epsfbox{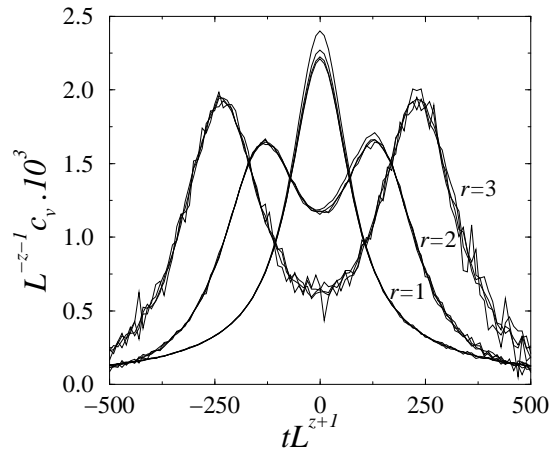}}
\end{center}
\caption{\label{fig3} Scaling plot of the specific heat of
the quantum Potts chain in the large-$q$ limit for the regular 
($r=1$) and the Fibonacci lattice ($r=2,3$). The lines are for 
finite 
systems with sizes $L=89,~233,~610$, and $1597$. For the 
Fibonacci 
lattice 
a double-peak structure is obtained.
} 
\end{figure}
%%%%%%%%%%%%%%%%%%%%%%%%%%%%%%%%%%%%%%%%%%%%%%%%%%

According to Eq.~(\ref{e0_scaling}), the specific heat of the 
system, $c_v\sim\partial^2 e_0/\partial t^2$, 
diverges with $L$ as $c_v\sim L^3$ at the transition point, a 
behavior which is compatible with Eq.~(\ref{chi_cv_scaling}) 
with $z=2$. Outside the transition point the scaling function
has been calculated numerically and is shown in 
Fig.~\ref{fig3}.

%%%%%%%%%%%%%%%%% FIGURE %%%%%%%%%%%%%%%%%%%%%%%%%
\begin{figure}[tbh]
\epsfxsize=7truecm
\begin{center}
\mbox{\epsfbox{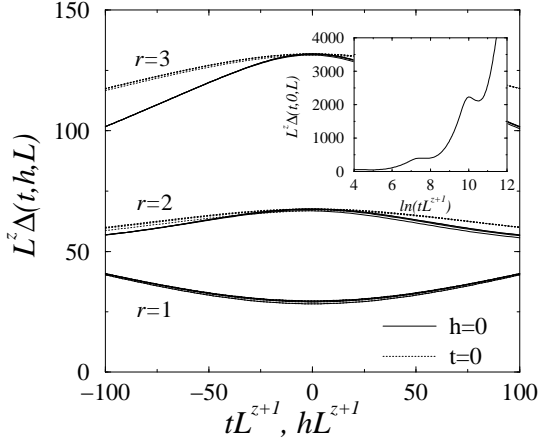}}
\end{center}
\caption{\label{fig4} Scaling plot of the energy 
gap of the quantum Potts chain in the large-$q$ 
limit for the regular ($r=1$) and
the Fibonacci lattice ($r=2,3$). The lines 
for finite systems with sizes $L=89,~233,~610$, 
and $1597$ indicate a good data collapse.
The inset shows log-periodic oscillations of the scaled gap at 
larger values of the scaling variable for $r=2$ and $L=89$. 
} 
\end{figure}
%%%%%%%%%%%%%%%%%%%%%%%%%%%%%%%%%%%%%%%%%%%%%%%%%%

Finally, we checked the scaling relation for the parallel 
correlation length in Eq.~(\ref{xi_scaling}). It 
is related to the scaling  behavior of the gap of 
the quantum Potts chain which is given by
\begin{equation}
\Delta(t,h,L)=L^{-z}\widetilde{\Delta}(tL^{z+1},hL^{z+1})\;.
\label{gap_scaling}
\end{equation} 
The results of numerical calculations for the scaling function 
$\widetilde{\Delta}(y,0)= \widetilde{\Delta}(0,-y)$ are shown 
in Fig.~\ref{fig4}. They agree with the analytical result 
$\widetilde{\Delta}(0,0)=3\pi^2$ [see Eq.~(21) in 
Ref.~\onlinecite{ic}].

\subsection{Fibonacci lattice}

The 1D Fibonacci lattice is composed of two units 
$A$ and $B$, and starting from $A$ the
sequence is generated through substitutions 
$A\to AB$ and $B\to A$. The couplings
or transverse fields on the lattice are also 
two valued, $\Gamma_A$ and $\Gamma_B$, depending on the
letter in the sequence at the given 
position. The strength of the quasiperiodic perturbation
is measured by the ratio $r=\Gamma_B/\Gamma_A$.
It has been noticed in
Refs.~\onlinecite{penrose1} and~\onlinecite{penrose2}, that the 
2D layered Fibonacci lattice and the 2D Penrose
quasilattice have a similar structure. On the 
Penrose lattice, starting from a
straightlike surface, two different types of 
parallel layers can be identified,
which are distributed according to the Fibonacci sequence. 
Consequently the SID phenomenon studied here on the layered 
Fibonacci lattice should behave in the same way 
on the Penrose lattice.

Considering the Potts model in the large-$q$ 
limit\cite{duality} the effective Hamiltonian in 
Eq.~(\ref{htilde})
at the transition point $t=h=0$ takes the form 
of a 1D quasiperiodic Schr\"odinger Hamiltonian,
the properties of which have been extensively 
studied.\cite{sch_fi} In particular, the scaling properties of 
the spectrum of $\widetilde{\cal H}$ at special points, 
such as at the edge of the spectrum, are 
exactly known.\cite{scal_fi} 
%%%%%%%%%%%%%%%%% FIGURE %%%%%%%%%%%%%%%%%%%%%%%%%
\begin{figure}[tbh]
\epsfxsize=7truecm
\begin{center}
\mbox{\epsfbox{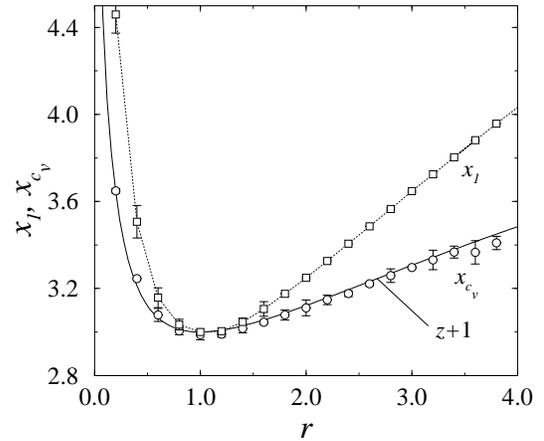}}
\end{center}
\caption{\label{fig5} Finite-size-scaling 
exponents of the specific heat $x_{c_v}$ and
surface magnetization $x_1$, as a function 
of the amplitude ratio $r$.
The solid line represents the exact scaling result 
$x_{c_v}=z+1$ with $z$ given
by Eqs.~(\protect\ref{sigma}) and (\protect\ref{z_sigma}). 
The dotted line is a guide to the eye.
} 
\end{figure}
%%%%%%%%%%%%%%%%%%%%%%%%%%%%%%%%%%%%%%%%%%%%%%%%%%

They are relevant
for the strongly anisotropic scaling 
in our problem.
The integrated density of states $D(\Delta E)$,
measured from the top of the spectrum of $-\widetilde{\cal H}$ 
has the scaling property\cite{scal_fi}
\begin{equation}
D(\Delta E)\sim\vert\Delta E\vert^{\sigma},\quad 
\Delta E \to 0 \;.
\label{DE_scaling}
\end{equation} 
Here
\begin{equation}
\frac{1}{\sigma}=\frac{\ln\lambda}{\ln\phi^2}\;,
\label{sigma}
\end{equation} 
where $\phi=(1+\sqrt{5})/2$ is the golden-mean ratio, 
$\lambda=\{8J-1+[(8J-1)^2-4]^{1/2}\}/2$ with
$J=[3+(25+16I)^{1/2}]/8$ and $I=(r+r^{-1})^2/4$.

A relation between the exponent $\sigma$ and the 
anisotropy exponent $z$ can be obtained
by noticing the scaling relation for the integrated density 
of states $D'=D(\Delta E')=b D(\Delta E)$
when lengths are rescaled as $L'=L/b$, 
since $D(\Delta E)$ is the number of states per unit 
length. Using also 
the scaling relation for the gaps $\Delta E'=
b^z \Delta E$, as given in Eq.~(\ref{gap_scaling}), 
we arrive at the result
\begin{equation}
z=\frac{1}{\sigma}\;.
\label{z_sigma}
\end{equation} 
Consequently, the anisotropy exponent of the Potts 
model on the Fibonacci lattice is exactly known. 
Thus we can check the bulk scaling relations derived for 
SID, which contains only the anisotropy exponent $z$ 
as a parameter. The scaling behavior of the energy density 
in Eq.~(\ref{e0_scaling}) with varying
temperature and field is illustrated in Fig.~\ref{fig1}.
An excellent data collapse is obtained for the lengths 
indicated, i.e., for Fibonacci numbers $F_n$ 
with an odd index $n$. A different scaling function 
is obtained for $n$ even. The scaling functions 
corresponding to the temperature and field
perturbations are no longer related through inversion,
as they were for the homogeneous problem. 
%%%%%%%%%%%%%%%%% FIGURE %%%%%%%%%%%%%%%%%%%%%%%%%
\begin{figure}[tbh]
\epsfxsize=7truecm
\begin{center}
\mbox{\epsfbox{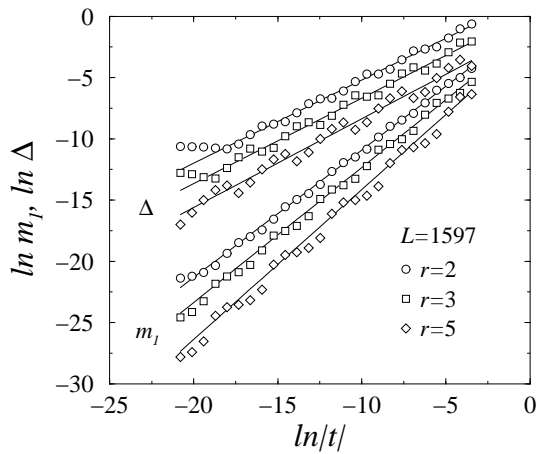}}
\end{center}
\caption{\label{fig6} Temperature dependence of 
the gap ($\Delta$) and the surface magnetization ($m_1$)
of the Potts model on the Fibonacci lattice 
with $L=1597$ for 
different values of $r$, on a
double-logarithmic scale. The
numerically calculated points, which show 
log-periodic oscillations, are compared to
asymptotic scaling results shown by straight 
lines. The slope of the lines are
$\nu_{\parallel}=z/(z+1)$ and $\beta_1=x_1/(z+1)$, 
for $\Delta$ and $m_1$, respectively,
where $x_1$ is taken from Fig.~\ref{fig5}.  
Deviations from the linear behavior at 
small $\vert t\vert$ values are due to finit-size effects.
Some sets of data were shifted vertically in order to improve
the legibility.}
\end{figure}
%%%%%%%%%%%%%%%%%%%%%%%%%%%%%%%%%%%%%%%%%%%%%%%%%%

We have also studied the finite-size scaling behavior of 
the specific heat at the transition point which is given by 
$c_v(t=0,h=0,L)\sim L^{x_{c_v}}$ and, according to
Eq.~(\ref{chi_cv_scaling}), $x_{c_v}=z+1$. The exponents
$x_{c_v}(r)$ were obtained through the extrapolation 
of two-point approximants using the 
Bulirsch-Stoer (BS) algorithm.\cite{bs,hs}
As shown in Fig.~\ref{fig5}
this scaling relation is indeed satisfied in the 
whole range of amplitude ratio $r$ investigated.

The scaling form of the specific heat in 
Eq.~(\ref{chi_cv_scaling}) is shown
in Fig.~\ref{fig3}. For the Fibonacci lattice the scaling
function around $t=0$ has a double peak structure 
with a distance between the peaks
increasing with $r$.

The finite-size scaling behavior of the energy gap 
given in Eq.~(\ref{gap_scaling}) is tested numerically 
in Fig.~\ref{fig4} as a function of temperature and field. 
The unexpected maximum of the scaled gap at 
the transition point can be attributed to the presence 
of log-periodic oscillations, as shown in the inset
for larger values of the scaling variable $tL^{z+1}$. 
Outside the transition point, the gap remains finite 
with an asymptotic temperature dependence 
$\Delta(t)\sim t^{\nu_{\parallel}}$,
with $\nu_{\parallel}=z/(z+1)$. The numerical results 
for the temperature dependence, which is
perturbed by log-periodic oscillations, is shown
in Fig.~\ref{fig6} for different values of $r$.

Next we present the results about the surface 
scaling properties of the Potts model on the Fibonacci 
lattice. The determination of the order parameter, even at the
surface, requires knowledge of the ground-state
eigenvector of $\widetilde{\cal H}$ as shown in 
Eq.~(\ref{magn_P}).
With lack of an exact knowledge of this eigenvector
we calculated numerically the surface magnetization 
at the transition point as a function of the size 
of the system,  and  the scaling exponent $x_1$,
defined in Eq.~(\ref{m1_scaling}), was obtained 
through the extrapolation of two-point approximants using the 
BS algorithm. It is shown in Fig.~\ref{fig5} 
as a function of $r$. The surface magnetization 
was also calculated as a function of $t$ for $L=1597$ 
in order to check the scaling relation
in Eq.~(\ref{beta_1}) for the exponent $\beta_1$.
The results are shown in Fig.~\ref{fig6}.

%%%%%%%%%%%%%%%%% FIGURE %%%%%%%%%%%%%%%%%%%%%%%%%
\begin{figure}[tbh]
\epsfxsize=7truecm
\begin{center}
\mbox{\epsfbox{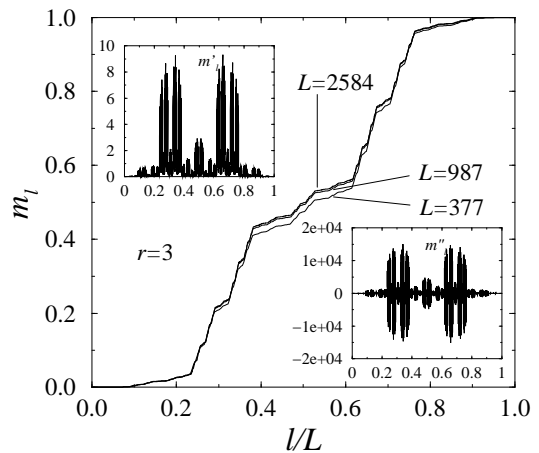}}
\end{center}
\caption{\label{fig7} Magnetization profile 
of the quantum Potts chain at the transition
point calculated on finite Fibonacci lattices 
with fixed- and free-boundary conditions.
In the insets the first and second finite differences 
are taken for $L=2584$.
} 
\end{figure}
%%%%%%%%%%%%%%%%%%%%%%%%%%%%%%%%%%%%%%%%%%%%%%%%%%

We close this section by presenting in 
Fig.~\ref{fig7} the magnetization profiles calculated
at the transition point on finite Fibonacci lattices. 
At first sight the shape of
the curves for different lengths looks similar;
however, a closer inspection
shows that with increasing size there are more 
and more points where a steplike behavior develops.
To look for a possible nonanalytical behavior 
of the curve in the thermodynamic
limit we have calculated the finite size differences 
$\Delta_1 m_l=L(m_{l+1}-m_{l-1})/2$ and
$\Delta_2 m_l=L^2(m_{l+1}+m_{l-1}-2 m_l)$. 
As seen in the insets of Fig.~\ref{fig7}, 
the derivatives show a self-similar, multifractal
structure. With increasing size $L$ the heights of 
the peaks are increasing, indicating
divergent first and second 
derivatives. Consequently $m_{l/L}$ is
an example of a physical observable with 
nonanalytical mathematical properties.

\section{Discussion}

In this paper we have presented a scaling picture 
for bulk first-order transitions associated with
SID, i.e., with a continuous surface 
transition. We have shown that fluctuations
of the interface separating the ordered (bulk) 
and disordered (surface) regions govern the scaling 
behavior of bulk quantities. Power-law 
singularities are present and the
bulk free-energy density shows a strongly 
anisotropic scaling behavior at the first-order transition 
point,
which can be associated with an anisotropic DFP.

The discontinuous nature of the bulk transition 
fixes the form of the scaling exponents,
which can all be expressed in terms of the 
anisotropy exponent $z$ and the dimension of the 
system. Also the functional form of the scaling 
functions is particular. For example, response functions at
the transition point have a power-law-type 
size-dependent peak, from which a $\delta$-function singularity
develops in the thermodynamic limit.

The scaling picture for SID has been tested on 
the 2D Potts model in the large-$q$ limit, both on
the regular lattice and on the layered Fibonacci 
lattice (the latter is related to the
2D Penrose quasilattice). For the Fibonacci 
lattice the anisotropy exponent $z$ varies continuously 
with the amplitude ratio of the aperiodicity in an exactly 
known 
way.
The study of the associated variation of
other scaling exponents provides us with a stronger test 
of the SID scaling relations.

The marginal nature of the quasiperiodic 
perturbation of the Fibonacci type should be 
explainable through a relevance-irrelevance criterion, like 
the Harris criterion\cite{har} for critical systems in the
presence of quenched disorder.

For second-order transitions the relevance or irrelevance 
of aperiodic perturbations is generally related
to the size dependence of the fluctuations of the sequence, 
which is measured by the wandering exponent $\omega$.
For a 1D variation, as considered in this paper, 
$\omega$ is defined through the cumulated deviation from 
the average\cite{quef}
\begin{equation}
\Delta_\Gamma(L_n)=\sum_{l=1}^{L_n}(\Gamma_l-\overline{\Gamma}) 
\sim L_n^{\omega}\;,
\label{dgamma}
\end{equation} 
where $\overline{\Gamma}=\lim_{L_n\to\infty} 
(1/L_n)\sum_{l=1}^{L_n}\Gamma_l$ is the average
field and $L_n$ is the length of the sequence 
after $n$ substitutions (i.e., the Fibonacci 
number $F_n$). As shown
in Refs.~\onlinecite{luck} and~\onlinecite{igloi94}, 
for a thermal aperiodic perturbation, the crossover 
exponent at the fixed point of 
the pure system is $\varphi=1+\nu_{\perp}(\omega-1)$,
where $\nu_{\perp}$ is the perpendicular correlation length 
exponent of the pure system.
Such a perturbation is relevant (irrelevant) 
when $\varphi>0$ ($\varphi<0$).	

A naive application of this criterion to our problem, 
with $\nu_{\perp}=1/3$ according to Eq.~(\ref{exp_hom}),
leads to a marginal value of the wandering exponent 
$\omega=-2$. Thus the perturbation should be relevant 
for the Fibonacci sequence with $\omega=-1$.

Actually, the self-dual form of the aperiodic perturbation
considered in this work, with $J_l=\Gamma_l$ 
in Eq.~(\ref{hamiltonop}), is weaker than a standard thermal 
perturbation for which, for example, $J_l$ fluctuates 
aperiodically while $\Gamma_l$ is kept constant.  

In order to obtain a relevance-irrelevance criterion
adapted to our problem, let us consider the form of the 
Hamiltonian 
matrix in Eq.~(\ref{htilde}) and compare the
perturbation associated with a deviation $t$ from 
the transition point, on the one hand, to the perturbation 
introduced by the aperiodic fluctuations on the other hand.

The temperature perturbation contributes to the diagonal 
elements
by terms of the form
\begin{equation}
t\sum_{j=1}^k\Gamma_j=\overline{\Gamma}tk
+t\sum_{j=1}^k(\Gamma_j-\overline{\Gamma})\,.
\label{dtemp}
\end{equation}
According to Eq.~(\ref{dgamma}) when $k$ is large the last term 
behaves as $tk^\omega$ 
and may be neglected since the wandering exponent 
is always smaller than~1. Taking the average at a 
length scale $L\gg1$ of the leading contribution
to Eq.~(\ref{dtemp}), one obtains the following estimate for 
the 
temperature
perturbation:
\begin{equation}
\delta_t(L)=\frac{\overline{\Gamma}t}{L}\sum_{k=1}^{L-1}k
\sim tL\;. 
\label{dtemp2}
\end{equation}
The aperiodic fluctuations contribute mainly to the 
off-diagonal 
terms 
and using Eq.~(\ref{dgamma}), their mean value at the length 
scale $L$ is given by
\begin{equation}
\delta_\Gamma(L)=\frac{1}{L}\Delta_\Gamma(L)\sim 
L^{\omega-1}\;.
\label{dgamma2}
\end{equation}
The appropriate length scale is the perpendicular correlation
length $\xi_\perp$ and the relevance of the aperiodic 
perturbation
depends on the value of the ratio
\begin{equation}
{\delta_\Gamma(\xi_\perp)\over\delta_t(\xi_\perp)}
\sim t^{1+\nu_\perp(\omega-2)}\;,
\label{ratio}
\end{equation}
where we used Eq.~(\ref{xi_perp}). Thus for self-dual 
aperiodic perturbations the crossover exponent is 
\begin{equation}
\varphi_{\rm sd}=1+\nu_\perp(\omega-2)\;.
\label{crossover}
\end{equation}
It follows that with $\nu_\perp=1/3$ such perturbations 
are marginal when $\omega=-1$, in agreement with our results
for the Fibonacci lattice.

Finally, let us briefly mention that we have also studied 
the effect of an aperiodic perturbation following 
the period-doubling sequence.\cite{col} This sequence is 
generated through the substitutions $A \to {AB}, {B} \to {AA}$,
and its wandering exponent is $\omega=0$. According 
to Eq.~(\ref{crossover}) this perturbation is relevant 
at the first-order transition point of the
$q$-state Potts model. Indeed, using the results of 
an exact renormalization group method in 
Ref.~\onlinecite{itksz},
the anisotropy exponent is found to be formally infinite. More
precisely the correlation lengths are 
related through
\begin{equation}
\ln\xi_{\parallel}\sim\xi_{\perp}^{\mu}\;,
\end{equation} 
where $\mu=1/2$ for the period-doubling sequence.\cite{dw}

\acknowledgments 
This work has been supported by the Hungarian
National Research Fund under Grant Nos. OTKA
TO23642, TO25139, TO34183, MO28418 and M36803, 
by the Ministry of Education under Grant
No. FKFP 87/2001 and by the Centre of 
Excellence Grant No. ICA1-CT-2000-70029.
The Laboratoire de Physique des Mat\'eriaux 
is Unit\'e Mixte de Recherche 
C.N.R.S. No. 7556.

\refrule

\end{multicols}


\begin{references}
\bibitem{rem} 
        Actually, in the large-$q$ limit, $e_0^{\rm             
        sing}$ is to be understood as the singular part of 
	$\sqrt{q}\langle0\vert\widetilde{\cal H}\vert0\rangle$
	according to Eq.~(\protect\ref{htilde}).
\bibitem{duality} 
        Since the sequence is not symmetric, the                
        Fibonacci Hamiltonian is only approximately self-dual.  
        Nevertheless, the critical point remains located at      
        $J_l=\Gamma_l$.
\bibitem{dw} Actually, the exact calculation of the exponent 
        $\mu$ 
	was done in Ref.~\protect\onlinecite{itksz} for 
	a directed walk problem. The aperiodic modulation 
	of the weights corresponds in the present problem to    
        the 
	two-letter substitutions obtained by replacing 
	$A$ by $AA$ and $B$ by $AB$ 
	in the period-doubling sequence. One expects 
	the same critical behavior since the wandering exponent 
	is not modified by this decoration of the sequence.
	This is supported by numerical data.
\end{references}
\end{document}